\documentclass[fleqn,twoside]{article}
\usepackage{espcrc2}
\usepackage{amssymb}

\title{%
\makebox[0pt][l]{\raisebox{4cm}[0pt][0pt]{%
\parbox{\textwidth}{\normalsize%
\mbox{}\hfill{DESY 03--142}\\
\mbox{}\hfill{Edinburgh 2003/14}\\
\mbox{}\hfill{LU-ITP 2003/018}\\
\mbox{}\hfill{ZIB-Report 03--30}
}}}%
Accelerating Hasenbusch's acceleration of Hybrid Monte
  Carlo\thanks{Poster presented by H.~St\"uben at Lattice 2003.}}

\author{%
A. Ali Khan\address{Institut f\"ur Physik, 
Humboldt-Universit\"at zu Berlin, 12489 Berlin, Germany},
T. Bakeyev\address{Joint Institute for Nuclear Research, 
141980 Dubna, Russia},
M. G\"ockeler\address{Institut f\"ur Theoretische Physik, 
Universit\"at Leipzig, 04109 Leipzig, Germany}$^{\mathrm{,}}$%
\address[UR]{Institut f\"ur 
Theoretische Physik, Universit\"at Regensburg, 93040 Regensburg, Germany},
R. Horsley\address{School of Physics, The University of Edinburgh, 
Edinburgh EH9 3JZ, UK},
D. Pleiter\address[NIC]{John von Neumann-Institut f\"ur Computing NIC, 
13738 Zeuthen, Germany},
P.E.L. Rakow\address{Theoretical Physics Division, 
Department of Mathematical Sciences, University of Liverpool, \\
Liverpool L69 3BX, UK},
A. Sch\"afer\addressmark[UR],\\
G. Schierholz\addressmark[NIC]$^{\mathrm{,}}$\address[DESY]{Deutsches
Elektronen-Synchrotron DESY, 22603 Hamburg, Germany}
and 
H. St\"uben\address{Konrad-Zuse-Zentrum f\"ur Informationstechnik Berlin, 
14195 Berlin, Germany}
(QCDSF Collaboration)
}

\newcommand{\Suv}{S_{\rm UV}}
\newcommand{\Sir}{S_{\rm IR}}
\newcommand{\Sdet}{S_{\rm det}}
\newcommand{\Nsteps}{N_{\rm steps}}
\newcommand{\Pacc}{P_{\rm acc}}
\newcommand{\Dgain}{D_{\rm gain}}
\newcommand{\Tint}{\tau_{\rm int}}
\newcommand{\Vuv}{V_{\rm UV}}
\newcommand{\Vir}{V_{\rm IR}}
\newcommand{\csw}{c_{\rm sw}}
\newcommand{\Nguess}{N_{\rm guess}}
\newcommand{\tCPU}{t_{\rm CPU}}

\newcommand{\be}{\begin{equation}}
\newcommand{\ee}{\end{equation}}
\newcommand{\bea}{\begin{eqnarray}}
\newcommand{\eea}{\end{eqnarray}}
\newcommand{\bc}{\begin{center}}
\newcommand{\ec}{\end{center}}
\newcommand{\bi}{\begin{itemize}}
\newcommand{\ei}{\end{itemize}}

\begin{document}

\begin{abstract}
Hasenbusch has proposed splitting the pseudo-fermionic action into two
parts, in order to speed-up Hybrid Monte Carlo simulations of QCD.  We
have tested a different splitting, also using clover-improved Wilson
fermions. An additional speed-up between 5 and 20\% over the original
proposal was achieved in production runs.
\vspace{.5pc}
\end{abstract}

\maketitle

\section{INTRODUCTION}

Hybrid Monte Carlo (HMC) \cite{HMC} is the standard algorithm employed
in numerical simulations of full QCD.  However, the computational cost
of such simulations grows rapidly with decreasing quark mass.  At
light quark mass (a) the condition number of the fermion matrix
increases, which leads to an increased number of iterations in solving
the corresponding system of linear equations, (b) the acceptance rate
decreases, which has to be compensated by decreasing the integration
step size, and (c) the autocorrelation time in units of trajectories
increases.

In \cite{Hasenbusch} Hasenbusch has proposed numerical methods to
improve conditions (a) and (b) in order to accelerate HMC simulations
with dynamical fermions.  He suggested splitting the fermion matrix
into two pieces both having a smaller condition number than the
original matrix.  For each factor a pseudo-fermionic field is
introduced and the Yang-Mills and fermionic parts of the action are
put onto different time-scales in the leap-frog integration.  These
methods were tested in simulations with clover-improved Wilson
fermions and a speed-up of 2 was obtained \cite{Hasenbusch+Jansen}.
The acceleration is greater at lower quark masses
\cite{Hasenbusch2003}.

The multiple time-scale approach was initially advocated in
\cite{Sexton+Weingarten} where Yang-Mills and pseudo-fermionic terms
were put onto different time-scales.  The idea was refined in
\cite{Peardon+Sexton} where the following criteria for an efficient
splitting of the action $S = \Suv + \Sir$ were formulated.  The force term
generated by $\Suv$ should be cheap to compute compared to $\Sir$.
And the splitting should mainly capture the high-frequency modes of the
system in $ \Suv $ and the low-frequency modes in $\Sir$.  In order to
achieve this, a low-order polynomial approximation for mimicking the
high-frequency modes was introduced in the fermionic action and the
action was split accordingly \cite{Peardon+Sexton}.

In this study the aforementioned methods are combined.  The fermion
matrix is split according to \cite{Hasenbusch}.  Following
\cite{Peardon+Sexton} the two fermionic contributions are put onto
different time-scales (this possibility was already mentioned in
\cite{Hasenbusch+Jansen} but no additional advantage was found).  In a
production run we compared our splitting with the splitting of
\cite{Hasenbusch+Jansen} and found an additional speed-up of about
20\% \cite{us}.  Here we report on the same comparison for a run at
smaller quark mass.

\section{NOTATION, TECHNICAL DETAILS}

\subsection{Actions}

We simulated two flavour QCD with clover-improved Wilson fermions
employing even/odd preconditioning.  The standard action for this
model reads
\be 
S_0[U,\phi^{\dag},\phi] = 
S_G[U] + \Sdet[U] + \phi^\dag (Q^\dag Q)^{-1} \phi \, \label{act0}
\ee
where
$S_G[U]$ is the standard Wilson plaquette action,
$\phi^\dag$ and $\phi$ are pseudo-fermion fields, and
\be 
\Sdet[U] = -2 \mbox{Tr} \log(1+T_{oo})\,,
\ee
\be 
Q = ({\bf 1} + T)_{ee} - M_{eo} ({\bf 1} + T)_{oo}^{-1} M_{oe}\,.
\ee
$T_{ee}$ $(T_{oo})$ is the clover matrix on even (odd) sites
\be
(T)_{a\alpha,b\beta}(x) =
{i \over 2}c_{sw}\kappa\sigma^{\alpha\beta}_{\mu\nu}
{\cal F}^{ab}_{\mu\nu}(x)\,.
\ee
$M_{eo}$ and $M_{oe}$ are Wilson hopping matrices connecting even with odd
and odd with even sites, respectively

The standard action is modified \cite{Hasenbusch} by introducing an
auxiliary matrix $W = Q + \rho$, $\rho \in \mathbb{R}$,
and pseudo-fermion fields $\chi^\dag$, $\chi$
\bea 
\lefteqn{S_1[U,\phi^{\dag},\phi,\chi^\dag,\chi] = S_G[U] + \Sdet[U]} \nonumber\\
& \mbox{} + 
\phi^\dag W(Q^\dag Q)^{-1} W^\dag \phi + \chi^\dag (W^\dag W)^{-1} \chi \,.
\label{act1}
\eea

\subsection{Multiple time-scales} \label{smult}

One step of the reversible integrator $V_n$ we used is given by
\cite{Peardon+Sexton}
\begin{eqnarray}
\lefteqn{V_n\left(\tau\right) = \Vir \left({\tau\over 2}\right) \times}
\label{integrator}
\\
& 
\makebox[5cm][l]{$\displaystyle
\left[\Vuv \left({\tau\over 2n}\right) 
       V_Q \left({\tau\over n}\right)
      \Vuv \left({\tau\over 2n}\right) \right]^n \times
      \Vir \left({\tau\over 2}\right)$}
\nonumber
\end{eqnarray}
where $n$ is a positive integer and the time-scales are $\tau$ and
$\tau/n$.  The effect of $V_Q$, $\Vuv$, $\Vir$ on the system
coordinates $\{ P,Q \}$ is:
\bea 
 V_Q(\tau) &:& Q \rightarrow Q + \tau P  \\
\Vuv(\tau) &:& P \rightarrow P - \tau \partial \Suv  \\
\Vir(\tau) &:& P \rightarrow P - \tau \partial \Sir 
\eea

\subsection{Splittings of the actions}

We performed simulations employing three splittings.  The first splitting is
based on $S_0$ (\ref{act0}). The other two are different splittings
of $S_1$ (\ref{act1}).

\medskip

\noindent
\textbf{Splitting A} (Sexton and Weingarten \cite{Sexton+Weingarten}):
\bea 
 \Suv & = & S_G[U] \nonumber \\
 \Sir & = & \Sdet[U] + \phi^\dag (Q^\dag Q)^{-1} \phi \label{zrta}
\eea

\noindent
\textbf{Splitting B} (Hasenbusch and Jansen \cite{Hasenbusch,Hasenbusch+Jansen}):
\bea
 \Suv & = & S_G[U] \nonumber \\
 \Sir & = & \Sdet[U] + 
 \phi^\dag W(Q^\dag Q)^{-1} W^\dag \phi \nonumber \\
 & & \mbox{} + \chi^\dag (W^\dag W)^{-1} \chi \label{zrtb}
\eea

\noindent
\textbf{Splitting C} (our proposal \cite{us}):
\bea 
 \Suv & = & S_G[U] + \Sdet[U] +\chi^\dag (W^\dag W)^{-1} \chi \nonumber \\
 \Sir & = & \phi^\dag W(Q^\dag Q)^{-1} W^\dag \phi
\label{zrt}
\eea

Our proposal (\ref{zrt}) is motivated by the hypothesis that most of the
high-frequency modes of the pseudo-fermion part of the action
(\ref{act1}) are located in $\chi^\dag (W^\dag W)^{-1} \chi$. We also
put the clover determinant $S_{det}[U]$ on the ``ultraviolet''
time-scale because the force generated by it is computationally
cheap. The computationally expensive term $\phi^\dag W(Q^\dag Q)^{-1}
W^\dag \phi$ is put on the ``infra-red'' time-scale.

\begin{table*}[t]
\caption{Parameters and statistics. (Statistics for each parameter set
  in Table 2.)}
\begin{tabular*}{\textwidth}{c%
@{\extracolsep{\fill}}c%
@{\extracolsep{\fill}}c%
@{\extracolsep{\fill}}c%
@{\extracolsep{\fill}}c%
@{\extracolsep{\fill}}c%
@{\extracolsep{\fill}}c%
@{\extracolsep{\fill}}c}
\hline
run & $V$ & $\beta$ & $\kappa$ & $\csw$ & $m_\pi / m_\rho$ & 
trajectory length & statistics \\
\hline
(I)  & $16^3 \times 32$ & 5.29 & 0.13550 & 1.9192 & $\mbox{} \approx 0.7$ &
1 & 300 trajectories\\
(II) & $24^3 \times 48$ & 5.25 & 0.13575 & 1.9603 & $\mbox{} \approx 0.6$ &
0.5 & 100 trajectories \\
\hline
\end{tabular*}
\end{table*}

\begin{table*}[t]
\caption{Further parameters and performance results. 
($\Nsteps$ is the number of integrator steps (\ref{integrator}).)}
\begin{tabular*}{\textwidth}{c%
@{\extracolsep{\fill}}c%
@{\extracolsep{\fill}}c%
@{\extracolsep{\fill}}c%
@{\extracolsep{\fill}}c%
@{\extracolsep{\fill}}c%
@{\extracolsep{\fill}}c%
@{\extracolsep{\fill}}c%
@{\extracolsep{\fill}}c%
@{\extracolsep{\fill}}c}
\hline
run & splitting & $\rho$ & $n$ & $\Nsteps$ &  $\Pacc$ & $N_Q$ & $N_W$ &
$N_Q+N_W$ & $\Dgain$\\
\hline
(I) & \textbf{A}  & 0   & 3 & 140 & 0.601 & 139492 & 0 & 139492 & 1 
\\[\smallskipamount]  
    & \textbf{B}  & 0.5 & 3 & 100 & 0.599 & 65951 & 5233 & 71184  & 1.95  \\ 
    &             & 0.2 & 3 & 70 & 0.664 & 47214  & 7378 & 54592 & 2.82 
\\[\smallskipamount]  
    & \textbf{C}  & 0.5 & 3 &  50 & 0.547 &  45160 & 7687 & 52847 & 2.40 \\ 
    &             & 0.2 & 3 &  40 & 0.663 &  32659 & 12373 & 45032 & 3.42 
\\[\medskipamount]  

\hline
(II) & \textbf{A}  & 0   & 3 & 180 & 0.780 & 267363 &     0 & 267363 & 1 
\\[\smallskipamount]  
     & \textbf{B}  & 0.2 & 3 &  90 & 0.891 &  89517 &  3242 &  92759 & 3.29 \\
     &             & 0.1 & 3 &  90 & 0.871 &  66432 &  5786 &  72218 & 4.13
\\[\smallskipamount]  
     & \textbf{C}  & 0.2 & 3 &  50 & 0.799 &  74002 &  7967 &  81969 & 3.34 \\ 
     &             & 0.1 & 3 &  50 & 0.896 &  57018 & 13624 &  70642 & 4.35 \\ 
\hline
\end{tabular*}
\end{table*}

\subsection{Solver}

The standard \emph{conjugate gradient} algorithm was used.  Starting
vectors were obtained from chronological inversion \cite{MRE} with
$\Nguess = 7$.  We checked reversibility by forward and backward
integration starting with thermalised configurations, whereupon
deviations of energies were less than $10^{-10}$.

\subsection{Computational gain}

\newcommand{\A}{\mbox{\scriptsize A}}
\newcommand{\B}{\mbox{\scriptsize B}}
\newcommand{\C}{\mbox{\scriptsize C}}

The CPU-cost is roughly given by $\tCPU \propto (N_Q + N_W)
\Tint$ where $N_Q$ and $N_W$ are the numbers of multiplications (per
trajectory) with $Q^\dagger Q$ and $W^\dagger W$, respectively.  In
order to estimate the computational gain we assume $\Tint \propto
1/\Pacc$ \cite{Hasenbusch+Jansen} and calculate the computational gain
of splittings B and C compared to A by
\be 
\Dgain^{(\B,\C)} = \frac{N_Q^{(\A)}}{N_Q^{(\B,\C)} + N_W^{(\B,\C)}}
\frac{\Pacc^{(\B,\C)}}{\Pacc^{(\A)}}\,.
\ee

\section{RESULTS}

We have tested splittings A, B and C in two production runs.  The
parameters of the runs are listed in Table~1.  Performance results are
shown in Table~2.  The values for run (I) are old results \cite{us}.
The values for run (II) are new.  
One sees that the speed-up is
considerable and that it grows with decreasing quark mass.  $\rho$ has
to be lowered at smaller quark masses.  In run (I) splitting C accelerates
the simulation by about 20\% better than splitting B.  In run (II) the
additional gain of using splitting C is only about 5\%.

In conclusion, the methods proposed by Hasenbusch work very well.  Our
variant of his method seems to perform even slightly better.  In both
cases the choice of the new parameter $\rho$ affects the speed-up
noticeably.  It would be interesting to know how the number of
integrator steps and the trajectory length influence the gain.

\section*{ACKNOWLEDGEMENTS} 

Computations were performed on the APEmille at NIC Zeuthen and on the
Hitachi SR8000-F1 at Leibniz-Rechenzentrum Munich.

\end{document}